\documentclass[aps,prb,reprint,twocolumn,superscriptaddress,floatfix,nofootinbib]{revtex4-1}
\usepackage{amsmath,amssymb,color,comment}
\usepackage[makeroom]{cancel}
\usepackage[caption=false]{subfig}
\usepackage{mathrsfs}
\usepackage{graphicx}
\usepackage{subfig}
\usepackage[countmax]{subfloat}
\usepackage[english]{babel}
\usepackage[bookmarks=true,colorlinks,linkcolor=cyan,urlcolor=red,citecolor=red]{hyperref}
\DeclareMathOperator{\Tr}{Tr}

 \usepackage{dsfont}

\newcommand{\Int}{\int\limits}

\newcommand{\sign}[1]{\text{sgn}{#1}}

\def\d{\mathrm d}

\definecolor{mypink1}{rgb}{0.858, 0.188, 0.478}

\def\d{\mathrm d}
\begin{document}

\title{Aging dynamics in quenched noisy long-range quantum Ising models}
\author{Jad C. Halimeh} 
\affiliation{Max Planck Institute for the Physics of Complex Systems, 01187 Dresden, Germany} 
\affiliation{Physics Department, Technical University of Munich, 85747 Garching, Germany}

\author{Matthias Punk}
\affiliation{Physics Department, Arnold Sommerfeld Center for Theoretical Physics,
and Center for NanoScience, Ludwig Maximilian University of Munich, 80333 Munich, Germany}

\author{Francesco Piazza}
\affiliation{Max Planck Institute for the Physics of Complex Systems, 01187 Dresden, Germany}

\begin{abstract}
We consider the $d$-dimensional transverse-field Ising model with power-law interactions $J/r^{d+\sigma}$ in the presence of a noisy longitudinal field with zero average. We study the longitudinal-magnetization dynamics of an initial paramagnetic state after a sudden switch-on of both the interactions and the noisy field. While the system eventually relaxes to an infinite-temperature state with vanishing magnetization correlations, we find that two-time correlation functions show aging at intermediate times. Moreover, for times shorter than the inverse noise strength $\kappa$ and distances longer than $a(J/\kappa)^{2/\sigma}$ with $a$ being the lattice spacing, we find a critical scaling regime of correlation and response functions consistent with the model A dynamical universality class with an initial-slip exponent $\theta=1$ and dynamical critical exponent $z=\sigma/2$. We obtain our results analytically by deriving an effective action for the magnetization field including the noise in a non-perturbative way. The above scaling regime is governed by a non-equilibrium fixed point dominated by the noise fluctuations.
\end{abstract}

\date{\today}
\maketitle

\section{Introduction}
The concept of universality is well established in closed classical
and quantum systems in equilibrium,\cite{Cardy_book,Sachdev_book} and
has been rigorously formulated through a number of advanced frameworks
such as scaling theory\cite{Cardy_book,Sachdev_book,Ma_book} and the
celebrated renormalization group
method.\cite{Wilson1971a,Wilson1971b,Wilson1972,Wilson1974,Fisher1974,Wilson1975}
Moreover, with the great degree of control currently available in
ion-trap\cite{Porras2004,Kim2009,Jurcevic2014} and ultracold-atom
setups,\cite{Levin_book,Yukalov2011,Bloch2008,Greiner2002}
the study of phase transitions and their effects on the nonequilibrium dynamics of closed quantum systems has
become experimentally possible.
%  and
% well-verified concept in many-body physics
%, where one is able to determine critical exponents in various models through the
%measurement of two-point correlators, for example.

In recent years, and in no small part due to this experimental
advancement, out-of-equilibrium criticality and dynamical phase
transitions have become the subject of extensive
theoretical\cite{Taeuber_book,Zvyagin2016,Heyl_review,Mori2017} and
experimental\cite{Nicklas2015,Zhang2017,Jurcevic2017,Neyenhuis2017,Flaeschner2018}
research. In equilibrium one is able to probe criticality
only in the ground state or thermal state of the system, whereas in
out-of-equilibrium systems there can be multiple instances of
criticality.\cite{hohenberg1977theory}
 More recently the attention has shifted to criticality in
prethermal states that
temporally precede the steady
state,\cite{berges_2004,Kitagawa2011,schmiedmaer_2012,Langen2013,kastner_2013,marino_2013,ciocchetta_2015,
chiocchetta_2017,Maraga2015,sciolla_2013} and to the long-time time-translation invariant
steady state itself.\cite{Sciolla2010,Sciolla2011,sciolla_2013} 
Even though the latter has been extensively studied and is known to be
connected to nonanalyticities in a dynamical analog of the free energy in
mean-field models,\cite{Homrighausen2017,Lang2018} the critical exponents involved in classifying the universality of the model are directly those known from equilibrium. 

Prethermal criticality, on the other hand, offers the unique possibility of studying truly out-of-equilibrium criticality, because in this case criticality is probed away from the steady state and the dynamics is not time-translation invariant. One of the most fascinating aspects of such prethermal criticality is the phenomenon of aging in systems quenched to a critical point \cite{calabrese_ageing_2006}. Aging occurs in the prethermal regime, before the system has relaxed into its steady state, and gives rise to a truly nonequilibrium critical exponent $\theta$ that can be extracted from the intermediate-time dynamics of the order parameter or the two-time $(s,t)$ correlation and response functions thereof. Due to the broken time-translation invariance the latter do not only depend on the time difference $t-s$, even at long times. In fact, the decay as a function of $t>s$ gets slower with larger $s$. In other words, the response of a system becomes slower with its \textit{waiting time} or \textit{age} $s$. This is the characteristic aging behavior shown also by structural glass and spin glasses, where even though the slow dynamics after a perturbation such as a quantum or temperature quench may be approaching an asymptotic value in single-time quantities, this does not mean that the system is approaching a stationary state.

Critical dynamics of thermal systems have been shown to exhibit such aging behavior in two-time correlation functions, indicating the absence of equilibration to a time-translation invariant stationary state.\cite{Janssen1989,gambassi_2011} This type of aging has been observed also in isolated systems described by $O(N)$ models,\cite{ciocchetta_2015,chiocchetta_2017,Maraga2015,sciolla_2013} where for critical quenches the response and correlation functions at small momenta exhibit time-dependence $\propto-t(s/t)^\theta$ and $(st)^{2-2\theta}$, respectively, for $t\gg s$.

Recently, the investigation has been extended to open systems. A coupling to (possibly non-thermal) baths might be present, together with other wanted or unwanted sources of environmental noise. As such, a theoretical framework describing how the openness of the system affects the aging behavior is desirable. Moreover, criticality can be fundamentally different between the closed and open system version of the same model, as is for instance the case for driven-dissipative systems even in the steady state\cite{sieberer_2013}. In the context of critical aging dynamics, dissipative systems, like $O(N)$ models in contact with a sub- or super-Ohmic bath\cite{schmalian_2014} or driven and lossy fully connected spin chains,\cite{lang_2016} as well as noisy models,\cite{marino_noisy_2012,kollath_2015} have been considered. In particular, aging has been predicted for lattice bosons in the presence of phase noise\cite{kollath_2015} and prethermalization for short-range Ising models with a noisy transverse field.\cite{marino_noisy_2012,lorenzo2017remnants}

In this work, we study the relaxation dynamics after a quench in a noisy spin system. We consider a transverse-field Ising model in $d$ spatial dimensions with power-law interactions $J/r^{d+\sigma}$ as function of distance $r$ in the presence of a longitudinal field with zero average and Gaussian Markovian fluctuations of strength $\kappa$. Starting from an initial paramagnetic state and suddenly switching on both the interactions and the longitudinal noise field, we compute the dynamics of the response and correlation functions of the longitudinal magnetization. The quantum Ising model with power-law interactions is a paradigmatic setup of condensed matter and quantum many-body physics. In addition to its simplicity, it encompasses infinitely many universality classes, including in just one dimension depending on the value of $\alpha$, and it has recently been the main protagonist in experiments on dynamical phase transitions.\cite{Zhang2017,Jurcevic2017,Flaeschner2018} As shown below, it also provides a quantum mechanical spin model where combined effects of noise and quench dynamics can be studied analytically, allowing our work to provide a formalism suitable for investigating criticality in current open many-body experiments.

\subsection{Summary of the main findings}

\emph{i).---} The two-point correlation functions for any given
space-time distance $t-s$ vanish exponentially as a function of the time $(t+s)/2$
elapsed after the quench, consistent with the fact that the system
reaches an infinite-temperature state.\cite{marino_noisy_2012} However, at intermediate times we find that 
correlation functions computed at two times $s$ and $t>s$ remain dependent on the ratio $s/t$ as is the case for aging systems (see Fig.~\ref{fig:plotsovert}). 
In particular, for $J\ll\kappa$ and times $t,s<1/\kappa$ we analytically obtain the response and correlation functions (for $t>s$) 
\begin{widetext}
\begin{align}\label{eq:DR11ShortTimes}
\langle\mathcal{M}_{\mathbf{p},t}\tilde{\mathcal{M}}_{\mathbf{p},s} \rangle\simeq&-\text{i}\cos[2\mathcal{R}_\mathbf{p}(t-s)]-\text{i}\frac{\kappa}{4\mathcal{R}_\mathbf{p}}\sin[2\mathcal{R}_\mathbf{p}(t-s)], \\
\langle \mathcal{M}_{\mathbf{p},t} \mathcal{M}_{-\mathbf{p},s}\rangle\simeq&\frac{4h^2}{\mathcal{R}_\mathbf{p}^2\left(4\mathcal{R}_\mathbf{p}^2+\kappa^2\right)}\Bigg\{2\mathcal{R}_\mathbf{p}\Big(2\mathcal{R}_\mathbf{p}\cos[2\mathcal{R}_\mathbf{p}(t-s)]-\kappa\sin[2\mathcal{R}_\mathbf{p}(t-s)]\Big)\text{e}^{-2\kappa s}-\left(4\mathcal{R}_\mathbf{p}^2+\kappa^2\right)\cos[2\mathcal{R}_\mathbf{p}(t-s)]\nonumber\\\label{eq:DK11ShortTimes}
&+\kappa\left(\kappa\cos\left[2\mathcal{R}_\mathbf{p}(t-s)\left(\frac{1+s/t}{1-s/t}\right)\right]+2\mathcal{R}_\mathbf{p}\sin\left[2\mathcal{R}_\mathbf{p}(t-s)\left(\frac{1+s/t}{1-s/t}\right)\right]\right)\Bigg\},
\end{align}
\end{widetext}
\noindent with $\mathcal{R}_\mathbf{p}=\sqrt{h\left(h-J_\mathbf{p}\right)}$, where $h$ is the transverse field strength and $J_{\mathbf{p}}$ is the
Fourier transform of the interaction profile $J/r^{d+\sigma}$. In~\eqref{eq:DR11ShortTimes} $\mathcal{M}$ is the longitudinal magnetization field and $\tilde{\mathcal{M}}$ is the
longitudinal response field. These two-time
functions show the breaking of time-translation
invariance.

\emph{ii).---} For a quench to the closed-system critical point $h=J_\mathbf{0}$ and for
  times $t,s\ll 1/\kappa$ as well for large distances such
that $|\mathbf{p}|a(J/\kappa)^{2/\sigma}\ll 1$ with $a$ the lattice spacing,
  the response and correlation functions enter an aging scaling
  regime consistent with the model A class\cite{hohenberg1977theory,calabrese_ageing_2006}
  \begin{align}\label{eq:DR_scaling}
&\langle\mathcal{M}_{\mathbf{p},t}\tilde{\mathcal{M}}_{\mathbf{p},s}
  \rangle\simeq \frac{\kappa}{J} (t-s)^{\frac{2-\upsilon-z}{z}}\left(\frac{t}{s}\right)^{\theta} \!\!\!\! F_R\!\left(|\mathbf{p}|(t-s)^{\frac{1}{z}},\frac{s}{t}\right),\\\label{eq:DK_scaling}
     &\langle \mathcal{M}_{\mathbf{p},t}
    \mathcal{M}_{\mathbf{-p},s}\rangle\simeq
    (t-s)^{\frac{2-\upsilon}{z}}\left(\frac{t}{s}\right)^{\theta-1} \!\!\!\! F_C\!\left(|\mathbf{p}|(t-s)^{\frac{1}{z}},\frac{s}{t}\right),
  \end{align}
  with 
\begin{align}
&F_R(x,y)=-\frac{\text{i}y}{4\sqrt{c_0c_\sigma}x^{\sigma/2}}\sin\left(2\sqrt{c_0c_\sigma}x^{\sigma/2}\right),\\
 &F_C(x,y)=4\frac{c_0}{c_\sigma}\frac{1}{x^\sigma}\nonumber\\
&\times\left[\cos\left(2\sqrt{c_0c_\sigma}x^{\sigma/2}\frac{1+y}{1-y}\right)-\cos\left(2\sqrt{c_0c_\sigma}x^{\sigma/2}\right)\right],
\end{align}
where we have expressed time in units of $1/J$ and space in units of $a$,
and with $c_{0,\sigma}$ being pure numbers.
We find the following critical exponents 
\begin{align}
\label{eq:critexp}
z=\frac{\sigma}{2},\;\theta=1,\; \upsilon=2-\sigma.
\end{align}
The dynamical critical exponent $z$ is the same as found in the closed
system.\cite{maghrebi2016causality}
  The above scaling regime is governed by a 
  non-equilibrium fixed point dominated by the noise fluctuations.
With the choice of our interaction
potential being $1/r^{d+\sigma}$, the above results should hold in arbitrary
dimensions in the weakly interacting regime $J\ll \kappa$.

\subsection{Organization of paper}
The rest of the paper is organized as follows. In Sec.~\ref{sec:formalism} we present our formalism, based on the Keldysh path-integral formulation for out-of-equilibrium many-body problems, that allows us to eventually derive a Langevin vector equation, from which two-time response and correlation functions of the longitudinal magnetization and its current can be extracted. In Sec.~\ref{sec:results} we present our numerical results for the two-time response and correlation functions of the longitudinal magnetization, and discuss the relaxation dynamics and aging observed therein, in addition to critical scaling behavior for quenches close to the critical point. We conclude in Sec.~\ref{sec:conclusion}, whilst providing further details of our derivation in Appendix~\ref{sec:App}.

\section{Formalism}
\label{sec:formalism}

Our goal is to derive an effective Langevin equation governing
the post-quench dynamics of the longitudinal magnetization in presence of a noisy
magnetic field. Within this semiclassical
approximation, the fluctuations in the magnetization are induced only
by the field noise. 
Upon formulating the
Martin-Siggia-Rose-De Dominicis-Janssen (MSRDJ) classical action\cite{kamenev_book}
corresponding to the above Langevin equation, we compute 
two-point correlators within a Gaussian approximation. 

\begin{figure}[]
\centering
\includegraphics [width=0.48\textwidth]{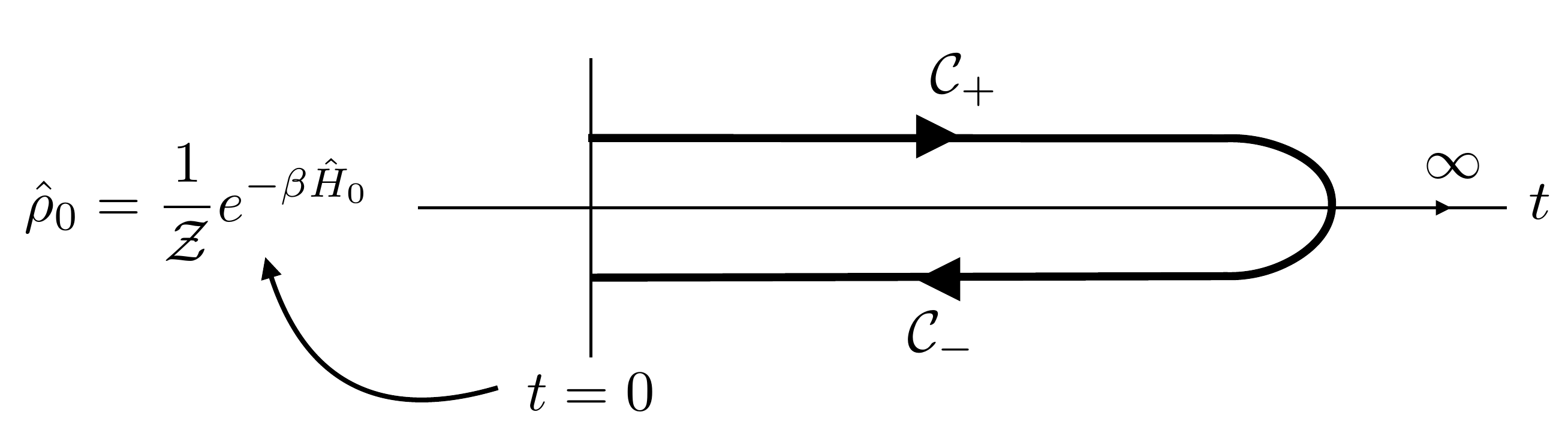}
\caption{Closed semi-infinite time contour employed in the path-integral formulation
  of the quench problem studied in this work. At the initial time $t=0$ our system is in a thermal state of the non-interacting noisless Hamiltonian
  $H_0$ given in~\eqref{eq:Hspin}.
}
\label{fig:contour}
\end{figure}

The starting point for the derivation of the Langevin equation is a
Hubbard-Stratonovich (HS) decoupling of the Ising interaction term
performed within a path-integral formulation of the quench
problem on the closed semi-infinite time contour\cite{danielewicz1984quantum} shown in Fig.~\ref{fig:contour}. The decoupling is performed
after mapping the spin-$1/2$ degrees of freedom to Schwinger bosons. In the absence of the noisy
field, the HS decoupling would be equivalent to the usual mean-field
decoupling. Here we instead include the noise non-perturbatively by solving
the Dyson equation for the two-point bosonic Green's function (GF) in a
self-consistent manner.

In Sec.~\ref{sec:model} we begin with introducing the quantum spin model used to describe
our system and its mapping to Schwinger bosons. In Sec.~\ref{sec:keldysh_approach} we then
describe the path-integral formulation of the problem on a closed semi-infinite time
contour. In Sec.~\ref{sec:MSR_action} we finally derive the
Langevin equation and its corresponding MSRDJ action.

\subsection{Model and Schwinger-boson mapping}
\label{sec:model}

We consider a transverse-field Ising model described by the Hamiltonian

\begin{align}\label{eq:Hspin}
\hat{H}&=\hat{H}_0+\hat{V},\\
\hat{H}_0&=-h\sum_i\hat{\sigma}^x_i,\\
\hat{V}&=-\frac{1}{2}\sum_{i\neq j}J_{ij}\hat{\sigma}^z_i\hat{\sigma}^z_j+\sum_i\eta_{i,t}\hat{\sigma}^z_i,
\end{align}
with $J_{ij}=J/|i-j|^{d+\sigma}$ the spin-spin coupling profile, $h$
the transverse-field strength, and $\hat{\sigma}^{x,z}$ the Pauli
matrices along the $x$ and $z$ directions, respectively. We add a
noisy longitudinal field $\eta_{i,t}$ with zero average and Gaussian fluctuations:
\begin{align}
\langle\eta_{i,t}\rangle&=0,\\
\langle\eta_{i,t}\eta_{j,t'}\rangle&=\frac{\kappa}{2}\delta_{i,j}\delta(t-t'),
\end{align}
with $\kappa$ a strength parameter.

We now use the Schwinger-boson representation of the Pauli spin operators,

\begin{align}\label{eq:SBmap1}
\hat{\sigma}^z&=\sign(\alpha)\hat{b}_{i,\alpha}^{\dagger}\hat{b}_{i,\alpha},\\
\label{eq:SBmap2}
\hat{\sigma}^x&=\hat{b}_{i,\alpha}^{\dagger}\hat{b}_{i,\bar{\alpha}},
\end{align}
with implied summation over spin indices, where $\hat{b}_{i\alpha}^{\left(\dagger\right)}$ is the Schwinger-boson annihilation (creation) operator on site $i$ for the two spin components $\alpha= \pm 1$ (or, equivalently in our notation, $\alpha=\uparrow\downarrow$) obeying the canonical commutation relations $[\hat{b}_{i,\alpha},\hat{b}_{j,\beta}^{\dagger}]=\delta_{i,j}\delta_{\alpha,\beta}$ and $[\hat{b}_{i,\alpha},\hat{b}_{j,\beta}]=0$. This allows us to express our Hamiltonian in the form

\begin{align}\nonumber
\hat{H}=&-\frac{1}{2}\sum_{i\neq j}J_{ij}\sign(\alpha)\sign(\beta)\hat{b}_{i,\alpha}^\dagger\hat{b}_{j,\beta}^\dagger\hat{b}_{i,\alpha}\hat{b}_{j,\beta}\\\nonumber
&-h\sum_i\hat{b}_{i\alpha}^{\dagger}\hat{b}_{i\bar{\alpha}}+\sum_i\eta_{i,t}\sign(\alpha)\hat{b}_{i\alpha}^{\dagger}\hat{b}_{i\alpha}\\\label{eq:H_SB}
&+\sum_i\lambda_{i,t}\big(\hat{b}_{i\alpha}^{\dagger}\hat{b}_{i\alpha}-2S\big),
\end{align}
where we have additionally included in the last line a Lagrange
multiplier $\lambda_{i,t}$ to constrain the number of bosons per site
to $2S$ at all times $t$, with $S=1/2$ the spin length in our model.

\subsection{Non-equilibrium path-integral formulation}
\label{sec:keldysh_approach}

We want to describe the dynamics within the following quench protocol.
At $t=0$ we prepare the system in a thermal state $\hat{\rho}_0=\text{e}^{-\beta \hat{H}_0}/\mathcal{Z}$ of the
non-interacting, noiseless Hamiltonian $\hat{H}_0$, and then let it subsequently
evolve with the full Hamiltonian~\eqref{eq:Hspin} in the presence of the
noise.

In order to compute non-equilibrium GFs we adopt a
path-integral formulation of the partition function on the closed semi-infinite time contour\cite{danielewicz1984quantum} $\mathcal{C}:0\to\infty\to0$ shown in Fig.~\ref{fig:contour}.

After HS-decoupling of the interaction term using the auxiliary
longitudinal magnetization field $\mathcal{M}$, 
the noise-averaged
partition function reads (see Appendix~\ref{sec:App} for more details)
\begin{align}
  \label{eq:Zhs}
\mathcal{Z}
=\Int\mathscr{D}[\bar{\phi}^\chi,\phi^\chi,\mathcal{M}^{\chi},\lambda,\eta]\text{e}^{\text{i}\mathcal{S}[\bar{\phi}^\chi,\phi^\chi, \mathcal{M}^{\chi},\lambda,\eta]},
\end{align}
with the action given by
\begin{widetext}
\begin{align}\nonumber
\mathcal{S}[\bar{\phi}^{\chi},\phi^{\chi},\mathcal{M}^{\chi},\lambda,\eta]=&\Int_{0}^{\infty}\d t\;\sum_{\chi=\pm}\chi\sum_i\left(\bar{\phi}_{i,\alpha,t}^{\chi}\text{i}\partial_t\phi_{i,\alpha,t}^{\chi}+h\bar{\phi}_{i,\alpha,t}^{\chi}\phi_{i,\bar{\alpha},t}^{\chi}-\eta_{i,t}\sign(\alpha)\bar{\phi}_{i,\alpha,t}^{\chi}\phi_{i,\alpha,t}^{\chi}-\lambda_{i,t}\bar{\phi}_{i,\alpha,t}^{\chi}\phi_{i,\alpha,t}^{\chi}\right)+\\\label{eq:actionforwardbackward}
&+\Int_{0}^{\infty}\d t\;\sum_{\chi=\pm}\chi\sum_{i\neq
  j}\left(\mathcal{M}^{\chi}_{j,t}J_{ij}\sign(\alpha)\bar{\phi}_{i,\alpha,t}^{\chi}\phi_{i,\alpha,t}^{\chi}-\mathcal{M}^{\chi}_{i,t}\frac{J_{ij}}{2}\mathcal{M}^{\chi}_{j,t}\right)+\frac{\text{i}}{\kappa}\Int_{0}^{\infty}\d
  t\;\sum_i\eta_{i,t}^2,
\end{align}
\end{widetext}
where $\chi=\pm$ indicates the forward ($0\to\infty$) or backward
($\infty\to0$) branch of the contour, respectively,
$\phi_{i,\alpha,t}^\chi$ are bosonic fields with
$\bar{\phi}_{i,\alpha,t}^\chi$ their complex conjugate, and
$\mathcal{M}_{i,t}^{\chi}$ is a real field. The noise field
$\eta_{i,t}$ is a classical field and therefore its value does not
depend on the contour branch.

\subsection{Langevin equation and MSRDJ action}
\label{sec:MSR_action}
\begin{figure}[]
\centering
\includegraphics [width=0.48\textwidth]{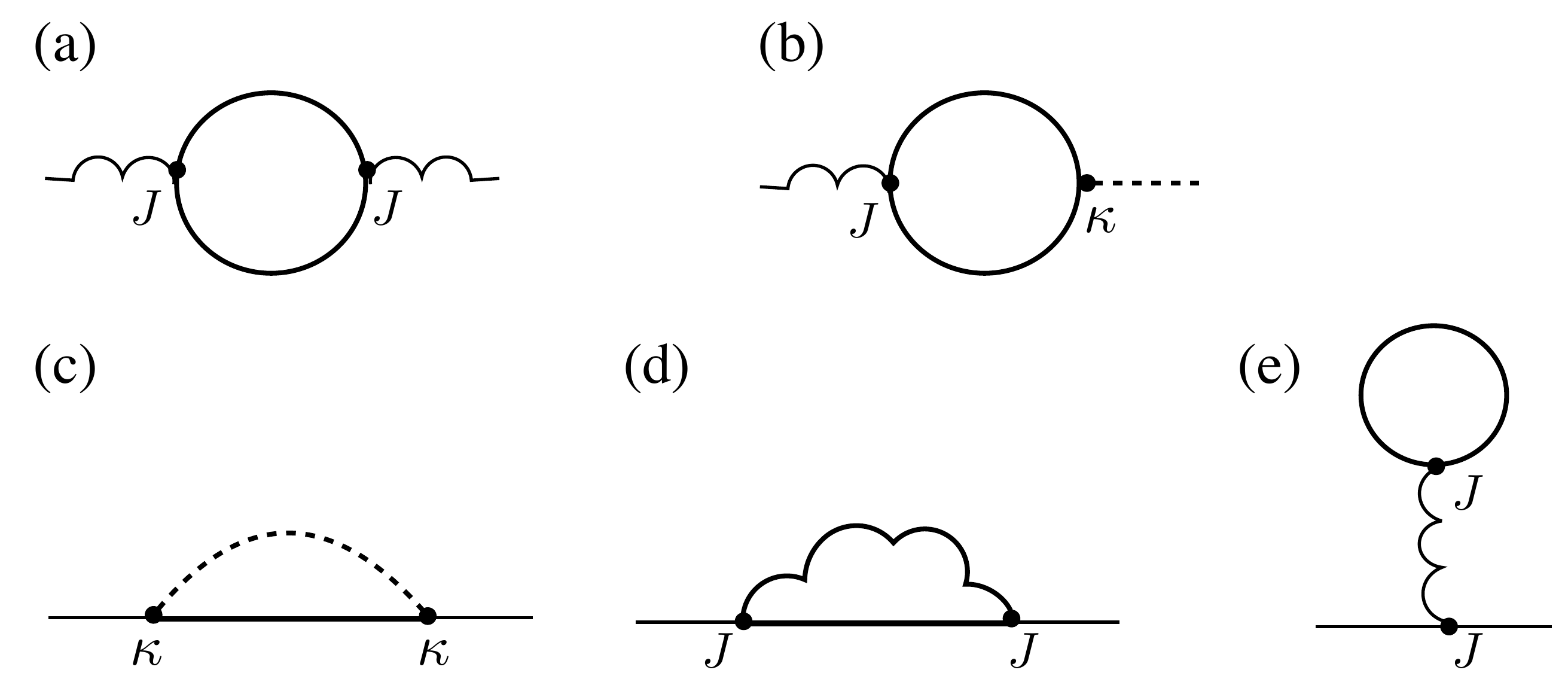}
\caption{Diagrammatic expression of the Gaussian action for the
  magnetization field $\mathcal{M}$ (wiggly lines), leading to the Langevin
  equation \eqref{eq:SPS}. The dimensionless noise field $\eta/\kappa$ is indicated by
  a dashed line while the Schwinger boson by a solid line. The
  magnetization couples to the bosons with coupling constant
  $J_{ij}$ while the dimensionless noise field with
  coupling constant $\kappa$. The
  non-equilibrium Green's functions are computed on the semi-infinite
  real-time closed contour of Fig.~\ref{fig:contour}. Contour indices as well as bosonic spin
  indices are suppressed.
  The Gaussian action is defined by diagrams (a)
  plus (b), whereby the bosonic loops are computed using the Green's
  functions dressed by the self-energy insertions (c), (d), and (e).
}
\label{fig:diagrams}
\end{figure}

In order to obtain the Langevin equation describing the dynamics of
the magnetization field $\mathcal{M}_{i,t}$, we perform a Keldysh rotation in~\eqref{eq:actionforwardbackward}, integrate out the \textit{classical} and \textit{quantum} bosonic fields
$\phi^{\text{cl}(\text{q})}_{i,\alpha,t}=(\phi^+_{i,\alpha,t}\pm\phi^-_{i,\alpha,t})/\sqrt{2}$, and consider the saddle-point equation of motion
up to linear order in $\mathcal{M}_{i,t}$
(see Appendix~\ref{sec:App} for a detailed derivation),
which reads
\begin{align}
&\mathcal{M}_{i,t}=-2\text{i}\Int_{0}^{\infty}\d\tau \Big(\sum_lJ_{il}\mathcal{M}_{l,\tau}-\eta_{i,\tau}\Big)\times\nonumber\\
&\times\Big(\Re
  G_{(i,\tau),(i,t)}^{\text{K},\uparrow\uparrow}G_{(i,t),(i,\tau)}^{\text{R},\uparrow\uparrow}-\text{i}\Im
  G_{(i,\tau),(i,t)}^{\text{K},\uparrow\downarrow}G_{(i,t),(i,\tau)}^{\text{R},\downarrow\uparrow}\Big),\label{eq:SPS}
\end{align}
with the classical component of the magnetization field
$\mathcal{M}_{i,t}\equiv\mathcal{M}_{i,t}^{\text{cl}}=(\mathcal{M}^{+}_{i,t}+\mathcal{M}^{-}_{i,t})/2$
and we have assumed our system is in the paramagnetic phase so that
$G^{\uparrow\uparrow}=G^{\downarrow\downarrow}$
and
$G^{\uparrow\downarrow}=G^{\downarrow\uparrow}$,
with the bosonic GFs defined as
$\text{i}G_{(j,t),(j',t')}^{\text{R},\alpha\alpha'}=\theta(t-t')\langle[\hat{b}_{j,\alpha}(t),
\hat{b}_{j',\alpha'}^\dag (t')]\rangle$ and $\text{i}G_{(j,t),(j',t')}^{\text{K},\alpha\alpha'}=\langle\{\hat{b}_{j,\alpha}
(t),\hat{b}_{j',\alpha'}^\dag (t')\}\rangle$.

The action corresponding to the above Langevin equation is
diagrammatically expressed by the sum of the two self-energies given
in Fig.~\ref{fig:diagrams}(a) and (b). In a purely Gaussian
approximation the bosonic GFs appearing in~\eqref{eq:SPS} would be
the bare propagators obtained for $\kappa=J_{ij}=0$. The natural improvement over
this crude approximation is the self-consistent Hartree-Fock treatment
corresponding to the self-energy corrections expressed diagrammatically
in Fig.~\ref{fig:diagrams}(c)-(e), whereby the $\mathcal{M}$ GF in Fig.~\ref{fig:diagrams}(d) and (e)
includes the loop corrections from Fig.~\ref{fig:diagrams}(a) and (b). 

However, we will restrict here to a weakly interacting case
$J\ll\kappa$ where we can neglect the corrections Fig.~\ref{fig:diagrams}(d) and (e) to the
bosonic self-energies, so that the latter are purely determined by the
noisy magnetic field. Due to the self-consistent resummation,
the diagrams (d) and (e) of Fig.~\ref{fig:diagrams} cannot be in
general neglected based on the perturbative criterion
$J\ll\kappa$. However, as we shall show in what follows, the presence of the
noise limits the growth of the magnetization correlation functions so
that no self-consistent enhancement takes place even at the critical point.

Within the above approximation we obtain the following bosonic response (see
Appendix~\ref{sec:App} for a detailed derivation)
\begin{align}\nonumber
G_{(i,t),(j,t')}^{\text{R}(\text{A}),\uparrow\uparrow}=&\mp\text{i}\text{e}^{\mp\frac{\kappa}{4}(t-t')}\Theta[\pm
  (t-t')]\cos[h(t-t')] \delta_{i,j},\\
G_{(i,t),(j,t')}^{\text{R}(\text{A}),\uparrow\downarrow}=&\pm\text{e}^{\mp\frac{\kappa}{4}(t-t')}\Theta[\pm (t-t')]\sin[h(t-t')]  \delta_{i,j},\label{eq:GR_bose}
\end{align}
and correlation functions
\begin{align}\nonumber
&G_{(i,t),(j,t')}^{\text{K},\uparrow\uparrow}=\delta_{i,j}\Big\{\\\nonumber
&-2\text{i}\left[\text{e}^{\frac{\kappa}{4}(t'-t)}\Theta(t-t')+\text{e}^{\frac{\kappa}{4}(t-t')}\Theta(t'-t)\right]\cos[h(t-t')]\\\nonumber
&+\text{e}^{-\frac{\kappa}{4}(t+t')}\left[\text{e}^{-\frac{\kappa}{2}t'}\Theta(t-t')+\text{e}^{-\frac{\kappa}{2}t}\Theta(t'-t)\right]\sin[h(t-t')]\Big\},\\[1em]\nonumber
&G_{(i,t),(j,t')}^{\text{K},\uparrow\downarrow}=\delta_{i,j}\Big\{\\\nonumber
&-\text{e}^{-\frac{\kappa}{4}(t+t')}\left[\text{e}^{-\frac{\kappa}{2}t'}\Theta(t-t')+\text{e}^{-\frac{\kappa}{2}t}\Theta(t'-t)\right]\cos[h(t-t')]\\
&+2\left[\text{e}^{\frac{\kappa}{4}(t'-t)}\Theta(t-t')+\text{e}^{\frac{\kappa}{4}(t-t')}\Theta(t'-t)\right]\sin[h(t-t')]\Big\}, \label{eq:GK_bose}
\end{align}
where, for simplicity, we consider our initial paramagnetic state to be at zero temperature% , and where due to $\mathbb{Z}_2$ symmetry we have $G_{(i,t),(j,t')}^{\text{R},\text{A},\text{K},\uparrow\uparrow}=G_{(i,t),(j,t')}^{\text{R},\text{A},\text{K},\downarrow\downarrow}$ and $G_{(i,t),(j,t')}^{\text{R},\text{A},\text{K},\uparrow\downarrow}=G_{(i,t),(j,t')}^{\text{R},\text{A},\text{K},\downarrow\uparrow}$
.
As explained in the Appendix~\ref{sec:App}, the above correlation functions are obtained by solving the Dyson
equation with the noise-induced self-energy of Fig.~\ref{fig:diagrams}(c) in a
non-perturbative, self-consistent manner. That is, the bosonic GF appearing in the
self-energy is not the bare one. This leads to a first-order linear
partial differential equation which is the two-time extension of the
Fokker-Planck or master equation employed for the non-interacting
fermionic model of Ref.~\onlinecite{marino_noisy_2012}.

Substituting the GFs \eqref{eq:GR_bose} and \eqref{eq:GK_bose} in the
Langevin equation \eqref{eq:SPS}, taking a $t$-derivative twice, and
going to Fourier space we obtain the following Langevin equation
\begin{align}
\label{eq:langevinA}
\partial_t\mathrm{M}_{\mathbf{p},t}^{a}&=\mathcal{A}_{\mathbf{p},t}^{ab}\mathrm{M}_{\mathbf{p},t}^b+\mathcal{B}_{\mathbf{p},t}^{ab}\xi_{\mathbf{p},t}^b,
\end{align}
where summation over repeated indices is implied and we defined the
vectors
$\mathrm{M}_{\mathbf{p},t}=(\mathcal{M}_{\mathbf{p},t}, \partial_t
\mathcal{M}_{\mathbf{p},t}/h)^\intercal$, $\xi_{\mathbf{p},t}=(0, \eta_{\mathbf{p},t}/h)^\intercal$ and the matrices
\begingroup
\renewcommand{\arraystretch}{1.5}
\begin{align}
\mathcal{A}_{\mathbf{p},t}^{ab}&=
\begin{pmatrix}
	0 & h \\
	-4h-\frac{\kappa^2}{4h}+4J_\mathbf{p}\text{e}^{-\kappa t} & -\kappa
\end{pmatrix}_{ab},\\
\mathcal{B}_{\mathbf{p},t}^{ab}&=
\begin{pmatrix}
	0 & 0 \\
	0 & -4h\text{e}^{-\kappa t}
\end{pmatrix}_{ab}.
\end{align}
Here we have introduced the Fourier transform $J_{\mathbf{p}}$ of the interaction matrix $J_{ij}$.

The Langevin equation with a first-order time-derivative has to take a
vector form since the original equation~\eqref{eq:SPS} for
$\mathcal{M}_{\mathbf{p},t}$ is a second-order differential equation (see Appendix~\ref{sec:App}).
Note that the vectorial form of the Langevin equation \eqref{eq:langevinA} belongs to the
model A dynamical universality class, but with the peculiarity that the friction and noise
act directly onto the current $\partial_t
\mathcal{M}$ and not onto the magnetization $\mathcal{M}$
itself. Still, there is no conserved quantity here as is the case for
the model A class.

In order to compute response and correlation functions of the
magnetization from the Langevin equation \eqref{eq:langevinA} we adopt
the stardard MSRDJ method to obtain the following classical action

\begin{align}
\nonumber
\mathcal{S}_\text{MSRDJ}[\tilde{\mathrm{M}}, \mathrm{M}]=&\sum_\mathbf{p}^\text{B.z.}\!\!\int_{0}^\infty\!\!\!\!\!\!\d
                           t\bigg[\frac{4\kappa}{\text{i}h^2}\tilde{\mathrm{M}}_{\mathbf{p},t}^a\mathcal{B}_{\mathbf{p},t}^{ab}\mathcal{B}_{\mathbf{p},t}^{bc}\tilde{\mathrm{M}}_{\mathbf{p},t}^c\\
                           &-2 \tilde{\mathrm{M}}_{\mathbf{p},t}^a\left(1^{ab}\partial_t-\mathcal{A}_{\mathbf{p},t}^{ab}\right) \mathrm{M}_{\mathbf{p},t}^b\bigg],\label{eq:MSR_action}
\end{align}
with the response-field vector $\tilde{\mathrm{M}}_{\mathbf{p},t}=(\tilde{\mathcal{M}}_{\mathbf{p},t}, \partial_t
\tilde{\mathcal{M}}_{\mathbf{p},t}/h)^\intercal$. The above
MSRDJ action is quadratic in the magnetization
field. However, it does not result from a purely Gaussian
approximation since it includes the loop corrections shown in Fig.~\ref{fig:diagrams}.

The longitudinal
magnetization response 
\begin{align}\label{eq:DRdef}
\text{i}[\mathcal{D}^{\text{R}}
  _{(\mathbf{p},t),(\mathbf{p}',t')}]^{ab}=\delta_{\mathbf{p},\mathbf{p'}}\langle \mathrm{M}_{\mathbf{p},t}^a \tilde{\mathrm{M}}_{\mathbf{p},t'}^b\rangle
\end{align}
and correlation function
\begin{align}\label{eq:DKdef}
\text{i}[\mathcal{D}^{\text{K}}_
  {(\mathbf{p},t),(\mathbf{p}',t')}]^{ab}=\delta_{\mathbf{p},-\mathbf{p'}}\langle \mathrm{M}_{\mathbf{p},t}^a \mathrm{M}_{-\mathbf{p},t'}^b\rangle
\end{align}
can be directly computed by inverting the matrix-valued differential
operator
% \begingroup
% \renewcommand*{\arraystretch}{1.5}
\begin{align}\label{eq:diffoperator}
\mathcal{D}_{(\mathbf{p},t),(\mathbf{p}',t')}^{-1}=&
	\begin{pmatrix}
	0 & \left[\mathcal{D}_{(\mathbf{p},t),(\mathbf{p'},t')}^{-1}\right]^{\text{R}^{\dag}} \\
	\left[\mathcal{D}_{(\mathbf{p},t),(\mathbf{p'},t')}^{-1}\right]^{\text{R}} & \left[\mathcal{D}^{-1}_{(\mathbf{p},t),(\mathbf{p'},t')}\right]^{\text{K}}
	\end{pmatrix},\\\nonumber
[\mathcal{D}_{(\mathbf{p},t),(\mathbf{p}',t')}^{-1}]^{\text{R}}=&\delta_{\mathbf{p},\mathbf{p'}}\delta(t-t')\\
&\times\begin{pmatrix}
	-\partial_t & h \\
	-4h-\frac{\kappa^2}{4h}+4J_\mathbf{p}\text{e}^{-\kappa t} & -\partial_t-\kappa
\end{pmatrix},\\
[\mathcal{D}_{(\mathbf{p},t),(\mathbf{p}',t')}^{-1}]^\text{K}=&\delta_{\mathbf{p},-\mathbf{p'}}\delta(t-t')
\begin{pmatrix}
	0 & 0 \\
	0 & -64\text{i}\kappa\text{e}^{-2\kappa t}
\end{pmatrix}.
\end{align}
% \endgroup

\section{Results}
\label{sec:results}

Starting from an initial paramagnetic state and suddenly
switching on both the interactions and the longitudinal noise field,
we want to study the dynamics of two-time response and correlation
functions of the longitudinal magnetization. We will first discuss the
typical behavior of response and correlation functions both in the
vicinity of and away from the critical point. We will subsequently turn
to the aging behavior at intermediate times, especially focusing on
what happens in the vicinity of the critical point. For simplicity, we have considered in our analysis a zero-temperature paramagnetic initial state, although our formalism can readily account for the finite-temperature case.

\begin{figure}[]
\centering
\includegraphics [width=0.48\textwidth]{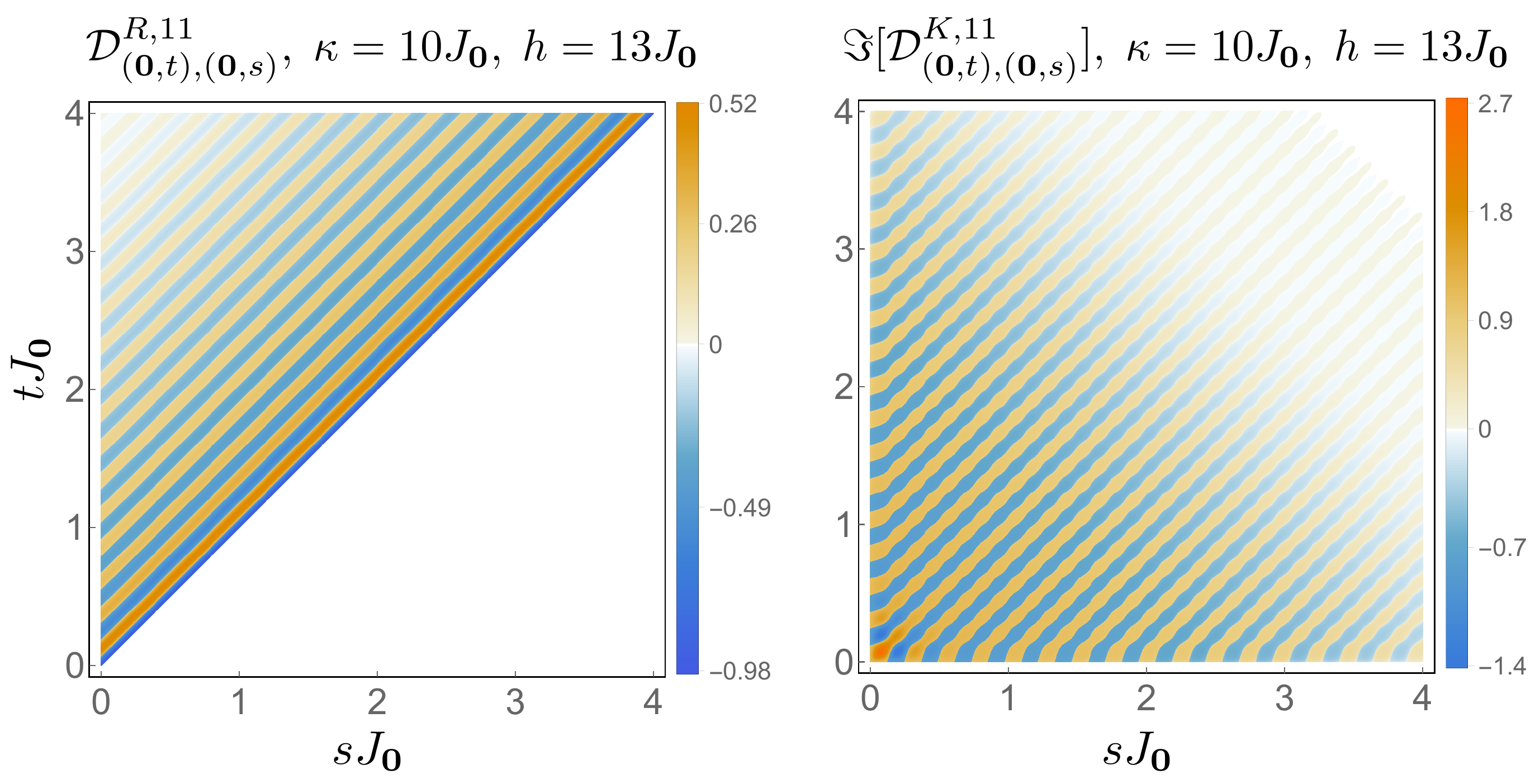}
\caption{(Color online). Response and correlation function of the longitudinal
  magnetization (see~\eqref{eq:DRdef} and~\eqref{eq:DKdef}) far away
from the critical point.}
\label{fig:2dplots_away}
\end{figure}

\begin{figure}[]
\centering
\includegraphics [width=0.48\textwidth]{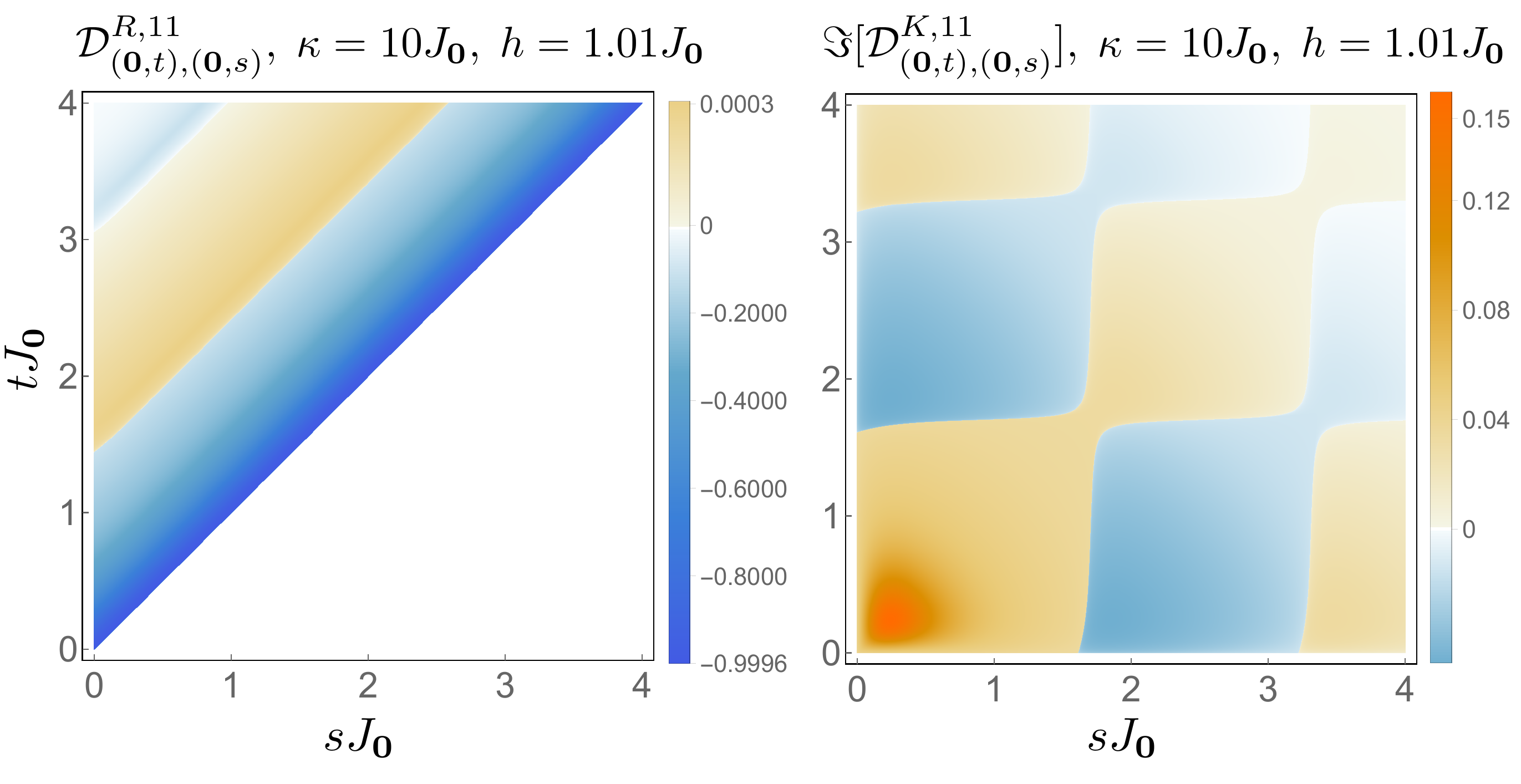}
\caption{(Color online). Response and correlation functions of the longitudinal
  magnetization (see~\eqref{eq:DRdef} and~\eqref{eq:DKdef}) close to the critical point.}
\label{fig:2dplots_close}
\end{figure}

\begin{figure*}[]
\centering
\includegraphics [width=0.83\textwidth]{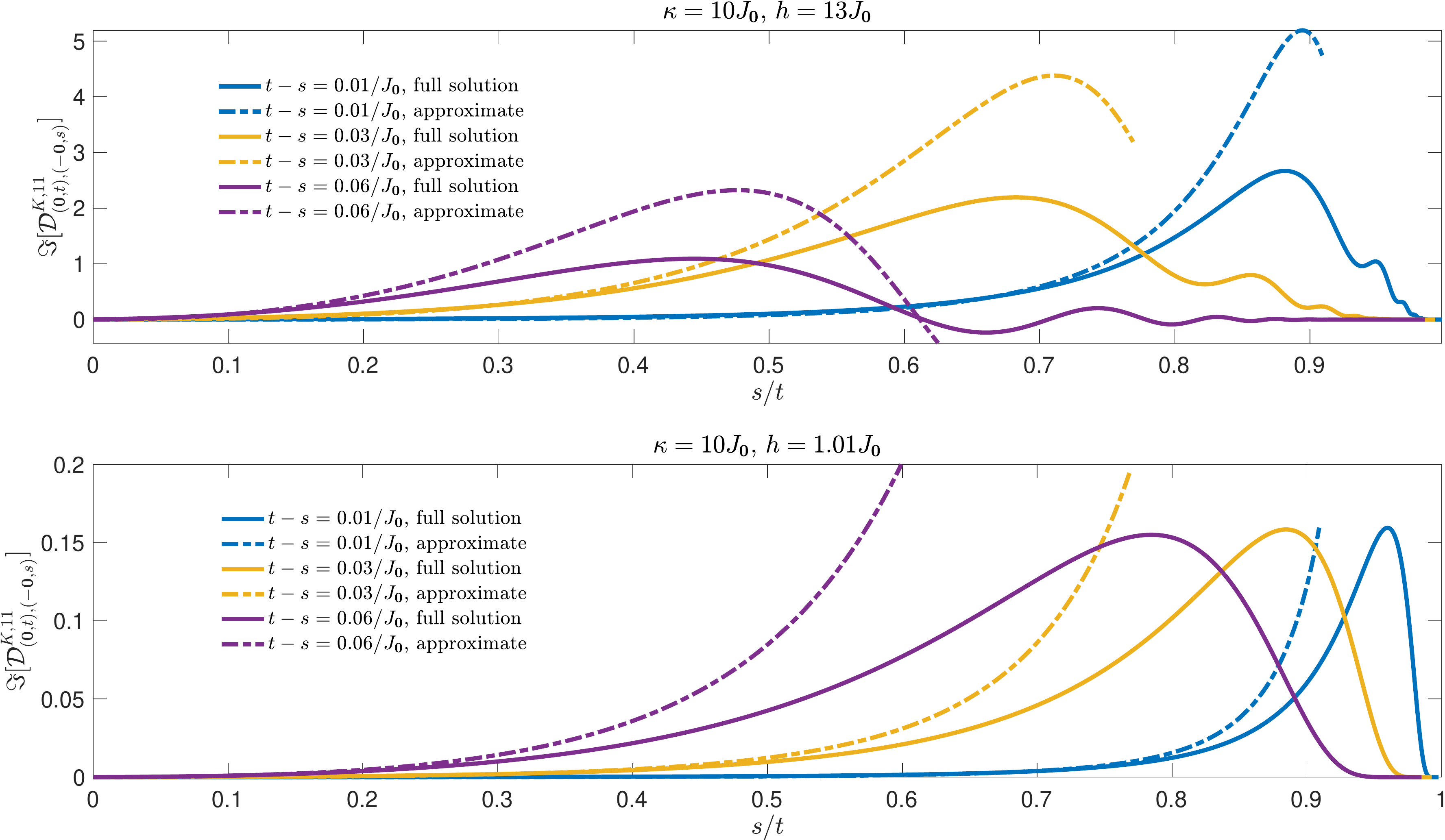}
\caption{(Color online). Behavior of the correlation of the longitudinal
  magnetization as a function of the ratio of the two times, both away
(upper panel) and close to (lower panel) the critical point. Different
colors correspond to different values of the time difference. The
dashed lines indicate the asymptotic solution at small times from~\eqref{eq:DK11ShortTimes} of the two-time correlation function.}
\label{fig:plotsovert}
\end{figure*}

\subsection{Relaxation dynamics and aging}

By inverting the operator \eqref{eq:diffoperator} we obtain the four
different response \eqref{eq:DRdef} and correlation \eqref{eq:DKdef}
functions. The four components correspond to the magnetization, the current,
and the two mixed correlators. In
Figs.~\ref{fig:2dplots_away} and~\ref{fig:2dplots_close} we provide an
example of the magnetization response and correlation
functions. Fig.~\ref{fig:2dplots_away} shows results computed far away
from the critical point of the closed system, which is defined by
$h=J_{\mathbf{0}}$. The response
function has the correct causal structure. It is also apparently
translation invariant, i.e.~it depends only on the time difference
$t-s$. The latter is a property of our approximation which
should be valid in the weakly interacting regime $\kappa\gg J$ (see
discussion in Sec.~\ref{sec:MSR_action}).
Before decaying exponentially at late times, both the response and
correlation functions show an oscillatory
behavior with frequency $2\mathcal{R}_{\mathbf{p}}$ with
$\mathcal{R}_\mathbf{p}=\sqrt{h\left(h-J_\mathbf{p}\right)}$
bounded from below by the distance $\mathcal{R}_\mathbf{0}$ to the critical point. When $\mathcal{R}_\mathbf{p}$ is small for $\mathbf{p}\to0$ and $h\to J_\mathbf{0}^+$, as is the case in Fig.~\ref{fig:2dplots_close}, the oscillations do not
have time to develop before the envelope decays exponentially to zero.

The exponential decay of the correlations towards zero for late
times is consistent with the expectation that the system relaxes to an
infinite-temperature state due to the presence of noise.\cite{marino_noisy_2012}
However, due to the quench that leads to the breaking of time-translation
invariance, two-point functions must still depend on both times $t,s$ (and
not only on their difference) up to a certain equilibration time
$\tau_{\rm eq}$. The
latter is set in our case by the exponential decay
and is proportional to the inverse noise strength: $\tau_{\rm eq}\sim
1/\kappa$. For times smaller than $1/\kappa$ we can analytically
compute response and correlation functions as given in~\eqref{eq:DR11ShortTimes} and \eqref{eq:DK11ShortTimes} for $t>s$.
They explicitly depend not only on the time difference $t-s$ but also
on their sum or, equivalently, on the ratio $s/t$. As long as these
functions depend on $s/t$ the system has not relaxed to its
equilibrium state, as is the case for systems showing aging. The
dependence of the correlation function on $s/t$ is shown in
Fig.~\ref{fig:plotsovert} for different values of the time
difference. After a given time the curves decay to zero and flatten
out, indicating relaxation to a time-translation invariant state. On
the other hand, for intermediate times there is a strong dependence on
$s/t$ and the correlation function agrees well with the analytic form
\eqref{eq:DK11ShortTimes}.

% The intermediate-time regime showing aging behavior can be arbitrarily
% extended towards late times by decreasing $\kappa$ and correspondingly
% $J$ so that the system remains in the weakly interacting regime
% $\kappa\gg J$. Most importantly, the limit $\kappa\rightarrow 0$ is
% then not smoothly connected to the noiseless case, as the
% fluctuations are governed by the longitudinal field noise and not by
% the quantum fluctuations originating from the Ising interactions in a
% transverse field. This feature is particularly relevant in the
% critical regime that we discuss next.

\subsection{Critical scaling behavior}

For a quench to the critical point of the noiseless system $h=J_\mathbf{0}$ and
restricting to $0<\sigma<2$ as well distances such that
$|\mathbf{p}|a\ll 1$, we
have 
% \[
% J_p \overset{p\to 0}{\simeq}\left\{\begin{array}{cc}J_0-J_\sigma|p|^\sigma & \sigma<2 \\
%     J_0-Jp^2\log|p| & \sigma=2 \\  J_0+J\zeta(\sigma-1) p^2 &
%                                                               \sigma>2\end{array}\right. ,
% \]
\begin{align}
J_\mathbf{p} /J \overset{\mathbf{p}\to \mathbf{0}}{\simeq}c_0-c_\sigma|\mathbf{p}|^\sigma,
\end{align}
% with $J_0=2J\zeta(1+\sigma)$, $J_\sigma=2J\cos(\pi\sigma/2)\Gamma[-\sigma]$ and 
where we have chosen the lattice spacing $a$ as our unit of length and
$c_0,c_\sigma$ are positive pure numbers depending on the dimension $d$. In
$d=1$ we have for instance $c_0=2\zeta(1+\sigma)$, $c_\sigma=-2\cos(\pi\sigma/2)\Gamma[-\sigma]$.
% For the sake of concreteness we will in the following restrict ourselves to the
% case $0<\sigma<2$ \jchcom{I believe this already covers all universality classes of the Ising model. If so, we may mention this.}.
At the critical point we have thus
\begin{align}
\mathcal{R}_\mathbf{p}\simeq J\sqrt{c_0 c_\sigma}|\mathbf{p}|^{\sigma/2}.
\end{align}

For times $t,s\ll 1/\kappa$ such that we can neglect the exponential
term in~\eqref{eq:DK11ShortTimes}, and for momenta such that $\kappa\gg
\mathcal{R}_\mathbf{p}$, i.e.~distances much longer than $a
(J/\kappa)^{2/\sigma}$, we can approximate the response and correlation
functions of~\eqref{eq:DR11ShortTimes} and~\eqref{eq:DK11ShortTimes}
as
\begin{align}\label{eq:DR11critical}
&\langle\mathcal{M}_{\mathbf{p},t}\tilde{\mathcal{M}}_{\mathbf{p},s}
  \rangle\simeq-\frac{\text{i}\kappa}{4 J \sqrt{c_0 c_\sigma}|\mathbf{p}|^{\sigma/2}}\sin[2 \sqrt{c_0 c_\sigma}|\mathbf{p}|^{\sigma/2} J(t-s)], \\\nonumber
&\langle \mathcal{M}_{\mathbf{p},t} \mathcal{M}_{-\mathbf{p},s}\rangle\simeq\bigg\{\cos\left[2 \sqrt{c_0 c_\sigma}|\mathbf{p}|^{\sigma/2} J(t-s)\left(\frac{1+s/t}{1-s/t}\right)\right]\\\label{eq:DK11critical}
&-\cos[2 \sqrt{c_0 c_\sigma}|\mathbf{p}|^{\sigma/2} J(t-s)]\bigg\}\frac{4c_0}{c_\sigma|\mathbf{p}|^\sigma}.
\end{align}
The above critical response and correlation functions can be brought
into the scaling form given in~\eqref{eq:DR_scaling} and~\eqref{eq:DK_scaling}, which is consistent with
the scaling form expected for the model A dynamical universality
class\cite{hohenberg1977theory,calabrese_ageing_2006} (we recall that our Langevin
equation takes the model A form \eqref{eq:langevinA} with no conserved
quantities). We thus get the
critical exponents given in~\eqref{eq:critexp}.
It is interesting to note that while the dynamical critical exponent $z$
agrees with the thermal equilibrium value, the initial-slip exponent
$\theta$ deviates from the value expected for the model A class in
contact with a thermal bath.\cite{calabrese_ageing_2006} This feature might be interpreted as an example of a ``hierarchical
shell structure of nonequilibrium criticality'' as proposed in Ref.~\onlinecite{sieberer_2014}, whereby the dynamical universality class is
further refined by an additional exponent in an open system without detailed balance.
In our case the universal exponents are valid in the
weakly-interacting regime $\kappa\gg J$, where the critical behavior is governed by a
fixed point dominated by the noisy longitudinal field with zero
average. Differently from Ref.~\onlinecite{sieberer_2014} where a
driven-dissipative steady state is considered, in our case detailed balance
at early times is broken by the post-quench aging behavior, whereby
the additional slip exponent emerges.

% We note that, according to the above scaling forms, except for the
% dimensionless factor $\kappa/J$ determining the overall size of the response
% function, there is no further dependence on any microscopic
% parameter. In particular, fixing the value of $\kappa/J\gg 1$ so that
% we are in the weakly interacting regime where our approach is valid, we
% can take the limit $\kappa\rightarrow 0$, which implies that the
% critical scaling region just described is not smoothly connected with
% the noiseless case. 

\section{Conclusion and outlook}
\label{sec:conclusion}

We considered the long-range transverse-field Ising model with
power-law interactions, where we perform a quench on the disordered
side of the equilibrium phase diagram in the presence of a noisy longitudinal
magnetic field with zero average. We showed that the dynamics
exhibits aging at short to intermediate times before the system
eventually settles into an infinite-temperature state. At these early
times and at long distances we also find a scaling regime governed by
a non-equilibrium fixed point dominated by the noise
fluctuations. Interestingly, the universal initial-slip exponent
$\theta=1$ that we find deviates from the value expected for the model A
dynamical universality class in contact with a thermal bath. This
suggests the emergence of a hierarchical shell structure of
nonequilibrium criticality in concomitance with aging in open
systems. We defer a thorough investigation of this scenario to a
future work.

An important feature of the present analysis is that it puts forward a new approach
for computing two-time response and correlation functions of quantum
spin models undergoing dephasing. 
The method involves the derivation of an effective Langevin equation
from which to compute two-point correlators within a self-consistent approximation. 
Our method can be readily extended to other quantum many-body systems with different symmetries under the influence of dephasing.

\section*{Acknowledgments}
The authors are grateful to Martin Eckstein, Michael Kastner, Johannes Lang, Mohammad Maghrebi, and Federico Tonielli for stimulating discussions, and to Jamir Marino for a careful reading of and valuable comments on our manuscript.

\appendix

\section{Further details on the derivation of the Langevin equation}
\label{sec:App}
\subsection{Effective Keldysh action}

In the main text, we presented the partition function
\begin{align}\nonumber
\mathcal{Z}% &\Int\mathscr{D}[\eta]\Int\mathscr{D}[\bar{\phi}^\chi,\phi^\chi,\lambda]\text{e}^{\text{i}\mathcal{S}_\eta}\text{e}^{\frac{1}{\kappa}\sum_i\Int_0^\infty\d t\;\eta_{i,t}^2}\\
  \label{eq:Z}
=\Int\mathscr{D}[\bar{\phi}^\chi,\phi^\chi,\lambda,\eta]\text{e}^{\text{i}\mathcal{S}[\bar{\phi}^\chi,\phi^\chi,\lambda,\eta]},
\end{align}
which one obtains upon averaging over the Gaussian noise, with the action given by
\begin{widetext}
\begin{align}
\mathcal{S}[\bar{\phi}^{\chi},\phi^{\chi},\lambda,\eta]=&\Int_{0}^{\infty}\d t\sum_{\chi=\pm}\chi\big(\sum_i\bar{\phi}_{i,\alpha,t}^{\chi}\text{i}\partial_t\phi_{i,\alpha,t}^{\chi}-H[\bar{\phi}^{\chi},\phi^{\chi},\lambda,\eta]\big)+\frac{\text{i}}{\kappa}\Int_{0}^{\infty}\d t\;\sum_i\eta_{i,t}^2,\\\nonumber
H[\bar{\phi}^{\chi},\phi^{\chi},\lambda,\eta]=&-\frac{1}{2}\sum_{i\neq j}J_{ij}\sign(\alpha)\sign(\beta)\bar{\phi}_{i,\alpha,t}^{\chi}\phi_{i,\alpha,t}^{\chi}\bar{\phi}_{j,\beta,t}^{\chi}\phi_{j,\beta,t}^{\chi}-h\sum_i\bar{\phi}_{i,\alpha,t}^{\chi}\phi_{i,\bar{\alpha},t}^{\chi}\\
&+\sum_i\eta_{i,t}\sign(\alpha)\bar{\phi}_{i,\alpha,t}^{\chi}\phi_{i,\alpha,t}^{\chi}+\sum_i\lambda_{i,t}\big(\bar{\phi}_{i,\alpha,t}^{\chi}\phi_{i,\alpha,t}^{\chi}-2S\big),
\end{align}
\end{widetext}
and $\chi=\pm$ indicates dynamics along the forward (backward) branch of the contour. We now perform the Hubbard-Stratonovich (HS) transformation by inserting in~\eqref{eq:Z} the ``fat unity''
\begin{align}
1=\Int\mathscr{D}[\mathcal{M}^{\chi}]\text{e}^{-\text{i}\Int_{0}^{\infty}\d t\;\sum_{i\neq j}\mathcal{M}_{i,t}^{\chi}\frac{\chi}{2}J_{ij}\mathcal{M}_{j,t}^{\chi}},
\end{align}
where a trivial prefactor has been absorbed into the measure. We shift the HS field $\mathcal{M}_{i,t}^{\chi}\to\mathcal{M}^{\chi}_{i,t}-\sign(\alpha)\bar{\phi}_{i,\alpha,t}^{\chi}\phi_{i,\alpha,t}^{\chi}$, which brings the Keldysh action in the form~\eqref{eq:actionforwardbackward}, and then perform the Keldysh rotation

\begin{align}\label{eq:Rot1}
&\phi_{i,\alpha,t}^{\chi}=\frac{1}{\sqrt{2}}\left(\phi_{i,\alpha,t}^{\text{cl}}+\chi\phi_{i,\alpha,t}^{\text{q}}\right),\\\label{eq:Rot2}
&\mathcal{M}_{i,t}^{\chi}=\mathcal{M}^\text{cl}_{i,t}+\chi\mathcal{M}^\text{q}_{i,t},
\end{align}
where ``cl'' and ``q'' denote the \textit{classical} and \textit{quantum} components of the field. Note that the fields $\eta_{i,t}$ and $\lambda_{i,t}$ are classical and, therefore, have no quantum component.~\eqref{eq:Rot1} and~\eqref{eq:Rot2} put the action~\eqref{eq:actionforwardbackward} in the form

\begin{align}\nonumber
&\mathcal{S}[\bar{\phi}^{\text{cl}(\text{q})},\phi^{\text{cl}(\text{q})},\lambda,\eta,\mathcal{M}^{\text{cl}(\text{q})}]=\\\nonumber
&\Int_{0}^{\infty}\d t\Int_{0}^{\infty}\d t'\;\sum_{i,j}\bar{\Phi}_{i,t}\left[\mathcal{G}^{-1}_{(i,t),(j,t')}+V_{(i,t),(j,t')}\right]\Phi_{j,t'}\\\label{eq:action}
&+\Int_{0}^{\infty}\d t\;\left[\frac{\text{i}}{\kappa}\sum_i\eta_{i,t}^2-\sum_{i\neq j}J_{ij}\left(\mathcal{M}^\text{cl}_{i,t}\mathcal{M}^\text{q}_{j,t}+\mathcal{M}^\text{q}_{i,t}\mathcal{M}^\text{cl}_{j,t}\right)\right],
\end{align}

\noindent where $\Phi_{i,t}=(\phi_{i,\uparrow,t}^{\text{cl}},\phi_{i,\downarrow,t}^{\text{cl}},\phi_{i,\uparrow,t}^{\text{q}},\phi_{i,\downarrow,t}^{\text{q}})^\intercal$, and we have split the bosonic-field part of the action into a free term described by

\begingroup
\renewcommand*{\arraystretch}{1.5}
\begin{align}
&\mathcal{G}_{(i,t),(j,t')}^{-1}=
	\begin{pmatrix}
	0 & \left[\mathcal{G}_{(i,t),(j,t')}^{-1}\right]^{\text{A}} \\
	\left[\mathcal{G}_{(i,t),(j,t')}^{-1}\right]^{\text{R}} & \left[\mathcal{G}_{(i,t),(j,t')}^{-1}\right]^{\text{K}}
	\end{pmatrix},
\end{align}
\endgroup
\begingroup
\renewcommand*{\arraystretch}{1.5}
\begin{align}
\label{eq:GRi}
&\left[\mathcal{G}_{(i,t),(j,t')}^{-1}\right]^{\text{R}(\text{A})}=\;\delta_{i,j}\delta(t-t')\left[(\text{i}\partial_t\pm\text{i}\varepsilon)\sigma^0+h\sigma^x\right],
\end{align}
\endgroup
\begingroup
\renewcommand*{\arraystretch}{1.5}
\begin{align}
\label{eq:GKi}
&\left[\mathcal{G}_{(i,t),(j,t')}^{-1}\right]^{\text{K}}=\;2\text{i}\varepsilon 
	\begin{pmatrix}
	F^{\uparrow\uparrow}(t,t') & F^{\uparrow\downarrow}(t,t') \\
	F^{\downarrow\uparrow}(t,t') & F^{\downarrow\downarrow}(t,t')
	\end{pmatrix}\delta_{i,j},
\end{align}
\endgroup
where~\eqref{eq:GKi} serves as a pure regularization term that is necessary for $\mathcal{G}^{-1}$ to be invertible with $\varepsilon\to0$ and $F^{\alpha\beta}(t,t')$ are (upon Wigner transformation) distribution functions, and into a ``source'' term described by 
\begingroup
\renewcommand{\arraystretch}{1.5}
\begin{align}
&V_{(i,t),(j,t')}=
	\begin{pmatrix}
	V_{(i,t),(j,t')}^{\text{q}} & V_{(i,t),(j,t')}^{\text{cl}} \\
	V_{(i,t),(j,t')}^{\text{cl}} & V_{(i,t),(j,t')}^{\text{q}}
	\end{pmatrix},\\
&V_{(i,t),(j,t')}^{\text{q}}=\;\sum_{l}J_{il}\mathcal{M}^\text{q}_{l,t}\sigma^z\delta_{i,j}\delta(t-t'),\\
&V_{(i,t),(j,t')}^{\text{cl}}=\;\delta_{i,j}\delta(t-t')\bigg[\big(\sum_{l}J_{il}\mathcal{M}^\text{cl}_{l,t}-\eta_{i,t}\big)\sigma^z-\lambda_{i,t}\sigma^0\bigg].
\end{align}
\endgroup
Inverting~\eqref{eq:GRi} yields the free retarded and advanced propagators
\begin{align}
\mathcal{G}_{(i,t),(j,t')}^{\text{R}(\text{A}),\uparrow\uparrow}&=\mp\text{i}\text{e}^{\mp\varepsilon(t-t')}\Theta[\pm (t-t')]\cos[h(t-t')],\\
\mathcal{G}_{(i,t),(j,t')}^{\text{R}(\text{A}),\uparrow\downarrow}&=\pm\text{e}^{\mp\varepsilon(t-t')}\Theta[\pm (t-t')]\sin[h(t-t')],
\end{align}
and where due to $\mathbb{Z}_2$ symmetry we have $\mathcal{G}_{(i,t),(j,t')}^{\text{R}(\text{A}),\uparrow\uparrow}=\mathcal{G}_{(i,t),(j,t')}^{\text{R}(\text{A},),\downarrow\downarrow}$ and $\mathcal{G}_{(i,t),(j,t')}^{\text{R}(\text{A}),\uparrow\downarrow}=\mathcal{G}_{(i,t),(j,t')}^{\text{R}(\text{A},),\downarrow\uparrow}$. Integrating out the bosonic degrees of freedom in the partition function

\begin{align}\nonumber
\mathcal{Z}=&\Int\mathscr{D}[\bar{\phi}^{\text{cl}(\text{q})},\phi^{\text{cl}(\text{q})},\lambda,\eta,\mathcal{M}^{\text{cl}(\text{q})}]\text{e}^{\text{i}\mathcal{S}[\bar{\phi}^{\text{cl}(\text{q})},\phi^{\text{cl}(\text{q})},\lambda,\eta,\mathcal{M}^{\text{cl}(\text{q})}]}\\
=&\Int\mathscr{D}[\lambda,\eta,\mathcal{M}^{\text{cl}(\text{q})}]\text{e}^{\text{i}\mathcal{S}_\text{eff}[\lambda,\eta,\mathcal{M}^{\text{cl}(\text{q})}]},
\end{align}
leads to the effective Keldysh action

\begin{align}\nonumber
&\mathcal{S}_{\text{eff}}[\lambda,\eta,\mathcal{M},\tilde{\mathcal{M}}]=\text{i}\Tr\ln\mathcal{G}^{-1}+\text{i}\Tr\ln\left(1+\mathcal{G}V\right)\\\label{eq:EffectiveAction}
&+\Int_{0}^{\infty}\d t\;\left[\frac{\text{i}}{\kappa}\sum_i\eta_{i,t}^2-\sum_{i\neq j}J_{ij}\left(\mathcal{M}^\text{cl}_{i,t}\mathcal{M}^\text{q}_{j,t}+\mathcal{M}^\text{q}_{i,t}\mathcal{M}^\text{cl}_{j,t}\right)\right].
\end{align}
$1+\mathcal{G}V$ is not diagonal in Keldysh, Nambu, or time space. We therefore Taylor-expand its inverse $(1+\mathcal{G}V)^{-1}$, which up to second order in $V$ leads to the saddle-point solution
\begin{align}
&\mathcal{M}_{i,t}=-2\text{i}\Int_{0}^{\infty}\d\tau \Big(\sum_lJ_{il}\mathcal{M}_{l,\tau}-\eta_{i,\tau}\Big)\times\nonumber\\
&\times\Big(\Re
  \mathcal{G}_{(i,\tau),(i,t)}^{\text{K},\uparrow\uparrow}\mathcal{G}_{(i,t),(i,\tau)}^{\text{R},\uparrow\uparrow}-\text{i}\Im
  \mathcal{G}_{(i,\tau),(i,t)}^{\text{K},\uparrow\downarrow}\mathcal{G}_{(i,t),(i,\tau)}^{\text{R},\downarrow\uparrow}\Big),\label{eq:SPS_noSC}
\end{align}
and this entails setting $\mathcal{M}^\text{q}\to0$, and we have thus dropped the superscript ``cl'' from the classical magnetization field $\mathcal{M}_{i,t}\equiv\mathcal{M}^\text{cl}_{i,t}$ for notational brevity.

% \begin{widetext}
% \begin{align}\label{eq:SPS}
% \mathcal{M}_{i,t}^{\text{cl}}=-2\text{i}\Int_{0}^{\infty}\d\tau\;\Big(\Re\mathcal{G}_{(i,\tau),(i,t)}^{\text{K},\uparrow\uparrow}\mathcal{G}_{(i,t),(i,\tau)}^{\text{R},\uparrow\uparrow}-\text{i}\Im\mathcal{G}_{(i,\tau),(i,t)}^{\text{K},\uparrow\downarrow}\mathcal{G}_{(i,t),(i,\tau)}^{\text{R},\downarrow\uparrow}\Big)\Big(\sum_lJ_{il}\mathcal{M}_{l,\tau}^{\text{cl}}-\eta_{i,\tau}\Big).
% \end{align}
% \end{widetext}

\subsection{Self-energies}
In order for $\mathcal{M}_{i,t}$ to be self-consistent, we must now calculate the self-energies. This is conveniently achieved by calculating the full propagator 

\begin{align}\nonumber
G^{uw}_{(l,\gamma,\tau),(m,\mu,\tau')}=&-\text{i}\langle\phi_{l,\gamma,\tau}^u\bar{\phi}_{m,\mu,\tau'}^w\rangle\\\label{eq:fullPropagator}
=&-\text{i}\langle\phi_{l,\gamma,\tau}^u\bar{\phi}_{m,\mu,\tau'}^w\text{e}^{\text{i}\mathcal{S}_V}\rangle_0,
\end{align}
by expanding up to second order in the interaction part of the action~\eqref{eq:action}

\begin{align}
\mathcal{S}_V[\bar{\phi},\phi,\lambda,\eta,\mathcal{M},\mathcal{M}^\text{q}]=&\Int_{0}^{\infty}\d t\Int_{0}^{\infty}\d t'\;\sum_{i,j}\bar{\Phi}_{i,t}V_{(i,t),(j,t')}\Phi_{j,t'}.
\end{align}
The self-energies arising from this expansion by approximating the Dyson equation as

\begin{align}
G=\mathcal{G}+\mathcal{G}\Sigma G\approx\mathcal{G}+\mathcal{G}\Sigma\mathcal{G},
\end{align}
are

\begin{align}\nonumber
\Sigma^{\text{R},\text{A},\text{K}}_{(i,t),(j,t')}=&\frac{\kappa}{2}\delta_{i,j}\delta(t-t')\sigma^0\\
&+\sum_{l,r}J_{il}J_{jr}\langle\mathcal{M}_{l,t}\mathcal{M}_{r,t'}\rangle_0\sigma^z\mathcal{G}^{\text{R},\text{A},\text{K}}_{(i,t),(j,t')}\sigma^z,
\end{align}
where, as mentioned previously, we take $\mathcal{M}^\text{q}\to0$. Recasting the Dyson equation in the form

\begingroup
\renewcommand*{\arraystretch}{1.5}
\begin{align}\label{eq:Dyson}
\mathds{1}_{4\times4}&=
\begin{pmatrix}
0 & \left[\mathcal{G}^{-1}\right]^{\text{A}}-\Sigma^{\text{A}} \\
\left[\mathcal{G}^{-1}\right]^{\text{R}}-\Sigma^{\text{R}} & -\Sigma^{\text{K}}
\end{pmatrix}
\begin{pmatrix}
G^{\text{K}} & G^{\text{R}} \\
G^{\text{A}} & 0
\end{pmatrix},
\end{align}
\endgroup
and taking $J\ll\kappa$ while considering the dynamics to always be restricted to the disordered phase, the retarded and advanced full propagators can then be calculated to be~\eqref{eq:GR_bose} in the main text.

Also from~\eqref{eq:Dyson}, upon enforcing self-consistency through replacing $\mathcal{G}^{\text{K}}$ with $G^{\text{K}}$ in the expression for $\Sigma^{\text{K}}$ (which for clarity we shall now call $\tilde{\Sigma}^{\text{K}}$) we obtain

\begin{align}
&\Big(\left[\mathcal{G}^{-1}\right]^{\text{R}}-\Sigma^{\text{R}}\Big)G^{\text{K}}=\tilde{\Sigma}^{\text{K}}G^{\text{A}},
\end{align}
from which we calculate the Keldysh full propagator~\eqref{eq:GK_bose} presented in the main text. Replacing $\mathcal{G}^{\text{K}(\text{R})}$ with $G^{\text{K}(\text{R})}$ in~\eqref{eq:SPS_noSC}, we arrive at~\eqref{eq:SPS} in the main text. Fourier-transforming from position into momentum space, and thereafter carrying out a time derivative twice, we arrive at the second-order differential equation

\begin{align}\nonumber
&\left(\partial^2_t+\kappa\partial_t+4h^2+\frac{\kappa^2}{4}-4hJ_\mathbf{p}\text{e}^{-\kappa t}\right)\mathcal{M}_{\mathbf{p},t}\\\label{eq:DiffEqM}
&=-4h\text{e}^{-\kappa t}\eta_{\mathbf{p},t},
\end{align}
which is then transformed into a first-order Langevin vector equation~\eqref{eq:langevinA} as illustrated in the main text.

\bibliography{DTFIM_biblio}

%--------------------------------------------------------
\end{document}